\documentclass[letterpaper, 10 pt, conference]{ieeeconf}  

\IEEEoverridecommandlockouts                              
\usepackage{graphicx} 
\usepackage{subcaption}
\graphicspath{{figures/}}
\usepackage{epstopdf}
\DeclareGraphicsExtensions{.png,.pdf}

\usepackage{amsmath} 
\usepackage{mathtools}

\DeclareMathOperator*{\argmin}{arg\,min}
\usepackage{amssymb}  
\usepackage{xcolor}

\newcommand{\bs}[1]{\boldsymbol{#1}}

\usepackage{array, multirow}
\newcolumntype{P}[1]{>{\centering\arraybackslash}p{#1}}

\usepackage{hhline}
\usepackage{geometry}
\title{\LARGE \bf
Sparsity Inducing System Representations for Policy Decompositions}

\author{Ashwin Khadke and Hartmut Geyer
\thanks{}
\thanks{Ashwin Khadke and Hartmut Geyer are with the Robotics Institute,
        Carnegie Mellon University, 5000 Forbes Avenue, Pittsburgh PA, 15213 USA
        {\tt\small \{akhadke, hgeyer\}@andrew.cmu.edu}}%
}

\begin{document}

\maketitle
\thispagestyle{empty}
\pagestyle{empty}

\begin{abstract}

Policy Decomposition (PoDec) is a framework that lessens the curse of dimensionality when deriving policies to optimal control problems. For a given system representation, i.e. the state variables and control inputs describing a system, PoDec generates strategies to decompose the joint optimization of policies for all control inputs.
Thereby, policies for different inputs are derived in a decoupled or cascaded fashion and as functions of some subsets of the state variables, leading to reduction in computation. However, the choice of system representation is crucial as it dictates the suboptimality of the resulting policies. We present a heuristic method to find a representation more amenable to decomposition. Our approach is based on the observation that every decomposition enforces a sparsity pattern in the resulting policies at the cost of optimality and a representation that already leads to a sparse optimal policy is likely to produce decompositions with lower suboptimalities.
As the optimal policy is not known we construct a system representation that sparsifies its LQR approximation. For a simplified biped, a 4 degree-of-freedom manipulator, and a quadcopter, we discover decompositions that offer $10\%$ reduction in trajectory costs over those identified by vanilla PoDec. Moreover, the decomposition policies produce trajectories with substantially lower costs compared to policies obtained from state-of-the-art reinforcement learning algorithms. 
\end{abstract}

\section{Introduction}
\begin{figure*}[ht!]
\centering
\begin{subfigure}{.35\textwidth}
  \centering
  \includegraphics[width=0.95\linewidth]{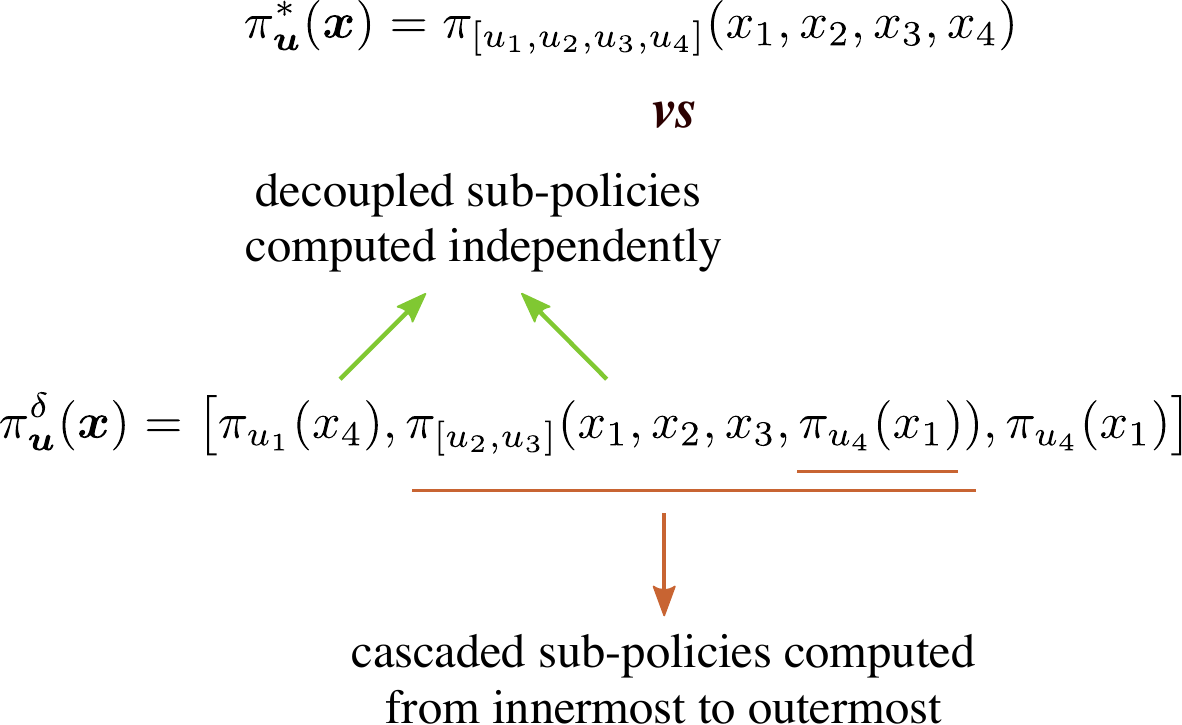}
  \caption{\scriptsize{An example decomposition for a  system with four states and four inputs.}}
  \label{f:podecidea}
\end{subfigure}%
\begin{subfigure}{.65\textwidth}
  \centering
  \includegraphics[width=1.0\linewidth]{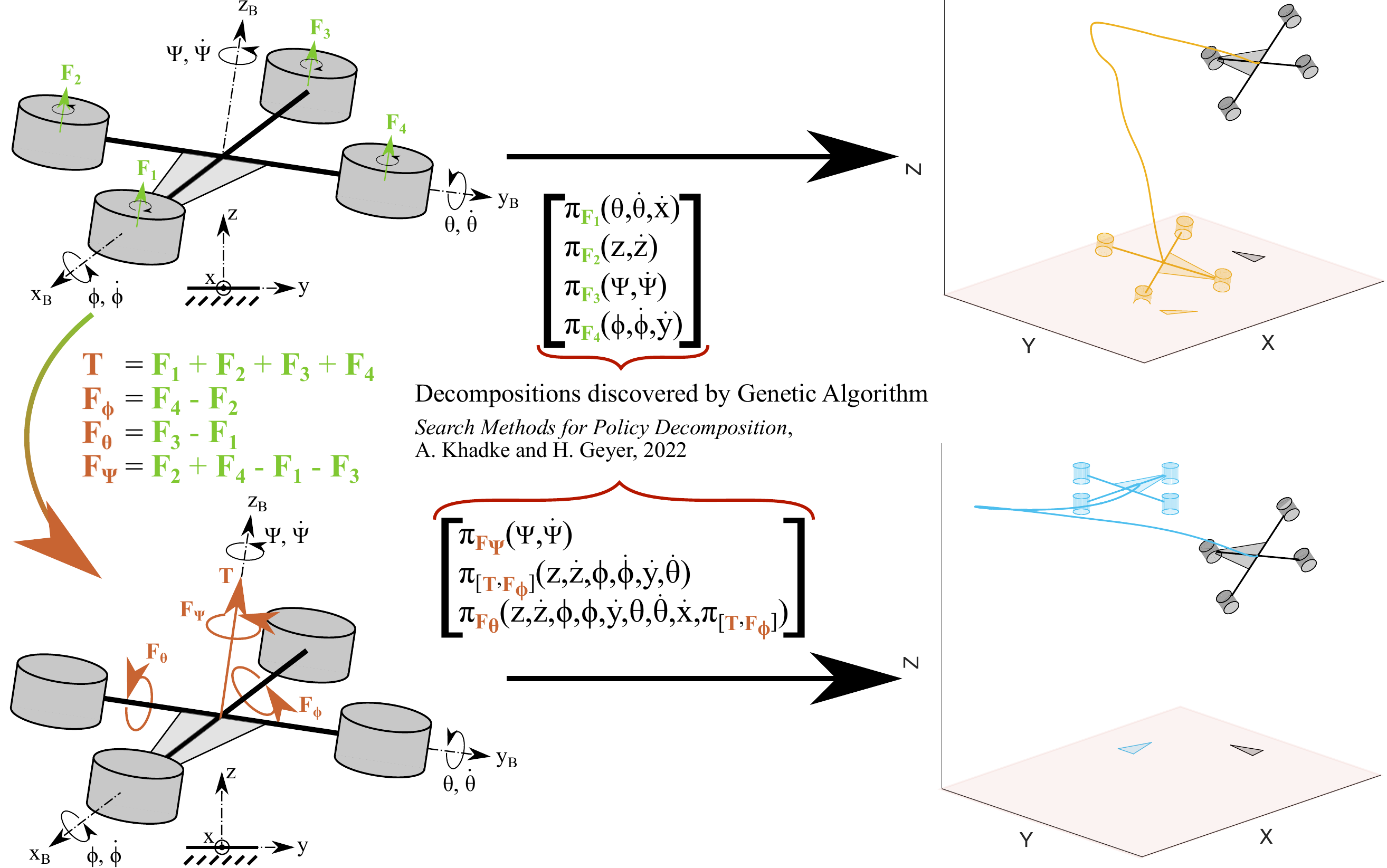}
  \caption{\scriptsize{Best decompositions for the two representations of the quadcopter system found using Genetic Algorithm \cite{Khadke:2022}.}}
  \label{f:choiceofrep}
\end{subfigure}
\caption{(\textbf{a}) PoDec uses the two fundamental strategies of cascading and decoupling to construct lower dimensional optimal control problems whose solutions are policies for different control inputs to the system. (\textbf{b}) Conduciveness to decomposition depends on the choice of system representation. The decomposition of policy computation for the quadcopter motor forces $F_i$ results in unstable closed-loop behavior. However, policy computation for the linearly transformed inputs $T, F_{\phi}, F_{\theta}, F_{\psi}$ is readily decomposable \cite{Khadke:2022}.}
\label{f:motivation}
\end{figure*}

Policy Decomposition (PoDec) \cite{Khadke:2021} is a framework to reduce computation in deriving global optimal control policies for complex nonlinear dynamical systems. PoDec achieves this reduction by constructing lower dimensional optimal control problems, allowing policies for different control inputs to be a function of only a subset of the state variables, and by computing these policies in a decoupled or a cascaded fashion (Fig.~\ref{f:podecidea}). 
For a given choice of state variables and control inputs describing the system, several strategies exist to decouple and cascade (or decompose) the policy computation. PoDec efficiently searches through the set of possible decompositions and identifies promising ones that balance suboptimality in closed-loop behavior with potential reduction in computation times \cite{Khadke:2022}. However, the choice of state variables and control inputs directly influences the quality of the resulting decompositions (Fig.~\ref{f:choiceofrep}).

Consider, as an example, a linear system with state variables $\bs{x} = [x_1, x_2]$, and control inputs $\bs{u} = [u_1, u_2]$,
\begin{equation}
    \begin{bmatrix}
    \dot{x}_1 \\
    \dot{x}_2
    \end{bmatrix} = \begin{bmatrix}
    x_2 \\ 
    x_1
    \end{bmatrix} + \begin{bmatrix}
    u_1 \\
    u_2
    \end{bmatrix}\nonumber
\end{equation}
The dynamics of this system are coupled and any decomposition, cascaded or decoupled, results in a suboptimal closed-loop control. However, dynamics of the linearly transformed system $\bs{y} = [y_1, y_2] = [x_1 + x_2, x_1 - x_2]$, $\bs{v} = [v_1, v_2] = [u_1 + u_2, u_1 - u_2]$ are decoupled,
\begin{equation}
    \begin{bmatrix}
    \dot{y}_1 \\
    \dot{y}_2
    \end{bmatrix} = \begin{bmatrix}
    y_1 \\ 
    -y_2
    \end{bmatrix} + \begin{bmatrix}
    v_1 \\
    v_2
    \end{bmatrix}\nonumber
\end{equation}
If the cost for the optimal control problem is separable for the $(\bs{y}, \bs{v})$ system, then policies for $v_1$ and $v_2$ can be derived independently as functions of $y_1$ and $y_2$ respectively, without inducing any suboptimality in the closed-loop behavior. 
For this simple linear system, by mere inspection, we identified a system representation that potentially generates better decompositions. But, it is difficult to find such representations for general nonlinear systems.

Several works have addressed the problem of finding compact state representations from raw sensory streams in a model-free MDP setting \cite{Anand:2019, Bellemare:2019, Gelada:2019, Dabney:2020}. We assume a known model of the system dynamics, and in such a setting system representations that simplify the policy computation are either hand designed \cite{Olfati:2001, Stilman:2005}, or derived based on some underlying network structure \cite{Sandberg:2009}, or by finding balanced realizations \cite{Moore:1981, Lall:2002, Gugercin:2004}, or by aggregating modes whose dynamics operate at similar time-scales \cite{Coderch:1983}. Alternatively, Koopman Operator based methods construct nonlinear features of the state variables such that dynamics in the feature-space can be approximated with a linear model \cite{Brunton:2016, Abraham:2017, Mamakoukas:2019}. 
However, all of these works are agnostic to the objective of the optimal control problem, and as such the system representations so identified may lead to poor closed-loop performance on decomposition. In contrast, our approach explicitly accounts for an approximation of the optimal policy and thus often produces more optimal decompositions.

This work extends the PoDec framework (reviewed in Section~\ref{sec:overview}) with an approach to find state and input mappings from a known system representation to a representation more amenable to decomposition. 
Every decomposition enforces sparsity in the resulting policy at the cost of optimality, as policies for different inputs are restricted to be only a function of some subset of the state variables, and a system representation that already leads to a sparse optimal policy is more likely to generate decompositions with lower suboptimalities.
Since the optimal policy is not known we compute its Linear Quadratic Regulator (LQR) based approximation, and derive linear state and input mappings that sparsify this approximation (Section~\ref{sec:transform}). We present extensive empirical evaluation of our approach with linear dynamical systems, which show that for $> 90\%$ instances the transformed representation offers decompositions with reduced suboptimalities (Section~\ref{subsec:linearemp}).
For the optimal control of three nonlinear dynamical systems: a simplified biped, a 4 degree-of-freedom planar manipulator and a quadcopter; the policies obtained by decomposing the transformed representation lead to $10\%$ reduction in closed-loop trajectory costs compared to those obtained from decomposing the original representation (Section~\ref{subsec:bipedmanipquad}). Finally, comparison with two state-of-the-art reinforcement learning methods, A2C \cite{Mnih:2016} and PPO \cite{Schulman:2017}, shows that decomposition policies produce trajectories with substantially lower costs (Section~\ref{subsec:bipedmanipquad}).

\section{Policy Decomposition}
\label{sec:overview}
For a general dynamical system with state $\bs{x}$ and input $\bs{u}$, 
\begin{equation} \label{eq:dynamics}
\dot{\bs{x}} = \bs{f}(\bs{x}, \boldsymbol{u})
\end{equation} 
the optimal control policy $\pi^*_{\bs{u}}(\bs{x})$ \emph{minimizes} an objective
\begin{equation} \label{eq:cost}
J_{} = \int_0^\infty e^{-\lambda t} c(\bs{x}(t), \bs{u}(t)) \: dt\nonumber
\end{equation}
that encodes the desired closed-loop behavior of the system. The objective is a discounted sum of costs accrued over time. We consider quadratic costs $c(\bs{x},\bs{u}) = (\bs{x} - \bs{x}^d)^T \bs{Q} (\bs{x} - \bs{x}^d) + (\bs{u}-\bs{u}^d)^T \bs{R} (\bs{u}-\bs{u}^d)$, where $\bs{x}^d$ is the goal state, and $\bs{u}^d$ the input that stabilizes the system at $\bs{x}^d$. Discount factor $\lambda$ quantifies the trade-off between immediate and future costs. 

Instead of jointly obtaining the optimal policies for all control inputs to a system $\pi^*_{\bs{u}}(\bs{x})$, PoDec computes sub-policies for individual subsets of inputs in a decoupled or cascaded fashion. These sub-policies are a function of only a subset of state variables and are derived by solving lower-dimensional optimal control problems leading to reduction in policy computation times. The sub-policies together form the decomposition policy $\pi^\delta_{\bs{u}}(\bs{x})$ for the entire system (Fig.~\ref{f:podecidea}). A sub-policy $\pi_{\bs{u}_i}(\bs{x}_i)$ is the optimal control policy for its corresponding subsystem, 
\begin{equation}
\dot{\bs{x}}_i = \bs{f}_i (\bs{x}_i, \bs{u}_i \mid\: \bar{\bs{x}}_i=\bar{\bs{x}}_i^d, \bar{\bs{u}}_i)\nonumber
\end{equation}
where $\bs{x}_i$ and $\bs{u}_i$ are subsets of $\bs{x}$ and $\bs{u}$, $\bs{f}_i$ only contains the dynamics associated with $\bs{x}_i$, and the complement state $\bar{\bs{x}}_i = \bs{x} \setminus \bs{x}_i$ is assumed to be constant. Complement inputs $\bar{\bs{u}}_i = \bs{u}\setminus\bs{u}_i$ that are decoupled from $\bs{u}_i$ are set to zero and sub-policies for those that are in cascade with $\bs{u}_i$ are used as is while computing $\pi_{\bs{u}_i}(\bs{x}_i)$. We represent the sub-policies as lookup tables over $\bs{x}_i$ and compute them using Policy Iteration \cite{Bertsekas:1995}, but any other representation and algorithm can be used.

Value-error, $\text{err}^\delta$, defined as the average difference over the state space $\mathcal{S}$ between the value functions $V^\delta$ and $V^*$ of the control policies obtained with and without decomposition, 
\begin{equation} \label{e:ValueError}
    \text{err}^\delta = \frac{1}{|\mathcal{S}|}\: \int_\mathcal{S} V^\delta(\boldsymbol{x}) - V^*(\boldsymbol{x}) \:\: d\boldsymbol{x}
\end{equation}
 quantifies the suboptimality of the policies obtained from decomposition $\delta$. $\text{err}^\delta$ cannot be computed without knowing $V^*$ and thus an estimate is derived by linearizing the dynamics about the goal state and input, 
 \begin{equation} \label{e:LQRdyn}
    \dot{\bs{x}} = \bs{A}(\bs{x} - \bs{x}^d) + \bs{B}(\bs{u} - \bs{u}^d)
\end{equation}
 \begin{equation}
\bs{A} = \left.\frac{\partial\bs{f}(\bs{x}, \bs{u})}{\partial\bs{x}}\right|_{(\bs{x}^d, \bs{u}^d)}, \:\:
\bs{B} = \left.\frac{\partial \bs{f}(\bs{x}, \bs{u})}{\partial \bs{u}}\right|_{(\bs{x}^d, \bs{u}^d)} \nonumber   
\end{equation} 
As the costs are quadratic the value function of the optimal policy for this linear system can be readily computed by solving the LQR problem \cite{Palanisamy:2015}, and we call it $V_{\text{lqr}}^*(\bs{x})$. The value-error estimate of decomposition $\delta$ is
\begin{equation} \label{e:LQREstimate}
\text{err}^\delta_\text{lqr} = \frac{1}{|\mathcal{S}|}\: \int_\mathcal{S} V^\delta_\text{lqr}(\boldsymbol{x}) - V^*_\text{lqr}(\boldsymbol{x}) \:\: d\boldsymbol{x}\nonumber
\end{equation}
where $V_{\text{lqr}}^{\delta}(\bs{x})$ is the value function for the equivalent decomposition of the linear system \cite{Khadke:2021}.

We find decompositions that optimally trade off reduction in computation with the suboptimality of the resulting policies using Genetic Algorithm as introduced in \cite{Khadke:2022}.

\section{Finding System Representations}
\label{sec:transform}
Consider the problem of obtaining an optimal controller for a quadcopter to hover in place (Fig.~\ref{f:choiceofrep}). Decomposition strategies for the policy optimization of motor thrusts $F_i$ that offer notable reduction in compute are highly suboptimal, and the best decomposition identified does not even produce stable closed-loop behavior. In contrast, suitable decompositions are found if one uses the linearly transformed inputs of net thrust $T$ and the roll, pitch and yaw differential thrusts ($F_{\phi}$, $F_{\theta}$ and $F_{\psi}$ respectively).
Here, we describe our approach to automatically discover such system representations, namely the choice of state variables and control inputs describing a system, that generate policy decompositions with reduced value-errors (Eq.~\ref{e:ValueError}). 

We pose the search for a system representation $(\bs{y},\bs{v})$ as the search for state and input mappings from a known representation $(\bs{x},\bs{u})$. For the purpose of this work we only consider linear and invertible mappings,
\begin{equation}
    \bs{y} = \bs{T}_y\;(\bs{x} - \bs{x}^d), \;\;\bs{v} = \bs{T}_v\;(\bs{u} - \bs{u}^d)\nonumber
\end{equation}
The dynamics of the system in  Eq.~(\ref{eq:dynamics}) can then be expressed in representation $(\bs{y}, \bs{v})$ as,
\begin{equation}
    \dot{\bs{y}} = \bs{T}_y\bs{f}(\bs{T}_y^{-1}\bs{y} + \bs{x}^d, \bs{T}_v^{-1}\bs{v} + \bs{u}^d)\nonumber
\end{equation}
Let $\delta_{(\bs{y}, \bs{v})}$ denote the decompositions of the system when expressed in the $(\bs{y}, \bs{v})$ representation. We desire mappings that minimize the value-error of the least suboptimal decomposition,
\begin{equation}
    \argmin_{\bs{T}_y, \bs{T}_v}\Big(\min_{\delta_{(\bs{y}, \bs{v})}}\text{err}^{\delta_{(\bs{y}, \bs{v})}}\Big)\label{e:transformcriteria}
\end{equation}
However, the above objective is not differentiable even for a linear system, and using gradient-free methods to search for mappings $\bs{T}_y$ and $\bs{T}_v$ along with the decomposition can be too computationally expensive. 
We tackle this problem with an alternative approach. Note that every decomposition induces sparsity in the resulting control policies $\pi^\delta$ at the cost of optimality. Therefore, a representation that already results in a sparse optimal policy is likely to result in lower suboptimality (or value-error) when decomposed. Finding mappings $\bs{T}_y$ and $\bs{T}_v$ to such a representation requires knowing the optimal policy. Subsequently, we obtain $\bs{T}_y$ and $\bs{T}_v$ based on an approximation of the optimal policy near the goal state $\bs{x}^d$. We linearize the system dynamics, 
\begin{equation}
    \dot{\bs{y}} = \bs{T}_y\bs{A}\bs{T}_y^{-1}\bs{y} + \bs{T}_y\bs{B}\bs{T}_v^{-1}\bs{v}\nonumber
\end{equation}
where $\bs{A}$ and $\bs{B}$ are defined in Eq.~\ref{e:LQRdyn}. Since our cost function is quadratic, the approximation to the optimal policy, $\pi^{\ast_{\text{approx}}}_{\bs{v}}(\bs{y})$, is the solution to the LQR problem
\begin{equation}
\begin{split}
    &\bs{A}^T\bs{P} + \bs{P}\bs{A} + \bs{Q} - \bs{P}\bs{B}\bs{R}^{-1}\bs{B}^T\bs{P} = \bs{0}\\&\;\;\;\;\;\;\;\;\;\;\;\;\;\;\;\;\;\;\pi^{\ast_{\text{approx}}}_{\bs{v}}(\bs{y}) = -\bs{\Theta}\bs{y}, \;\\&\text{where}\;\bs{\Theta} = \bs{T}_v\bs{K}^\ast\bs{T}_y^{-1},\;\bs{K}^\ast = \bs{R}^{-1}\bs{B}^T\bs{P}\label{e:lqrtransformed}
\end{split}
\end{equation}
We obtain a sparse parameterization to the approximation of the optimal policy using the singular value decomposition. Let $\bs{K}^\ast = \bs{U}_{\bs{K}^\ast}\bs{S}_{\bs{K}^\ast}\bs{V}_{\bs{K}^\ast}^T$ where $\bs{U}_{\bs{K}^\ast}\in\mathbb{R}^{m\times m}$ and $\bs{V}_{\bs{K}^\ast}\in\mathbb{R}^{n\times n}$ are orthogonal, then $\bs{T}_y = \bs{V}_{\bs{K}^\ast}^T$ and $\bs{T}_v = \bs{U}_{\bs{K}^\ast}^T$ result in a completely diagonal $\bs{\Theta} = \bs{S}_{\bs{K}^\ast}$.
Note that the singular value decomposition is not unique for rectangular matrices. We encounter wide $\bs{K}^\ast$s as all our systems have fewer control inputs than state variables $(m < n)$ and therefore we have some degree of freedom in choosing $\bs{U}_{\bs{K}^\ast}$ and $\bs{V}_{\bs{K}^\ast}$. Appendix \ref{apx:svd} presents a regularization strategy to resolve this non-uniqueness. 

\section{Results}
\label{sec:results}
First, we verify for linear systems if the singular value decomposition based mappings  indeed produce system representations that offer decompositions with lower value-errors.
Next, we investigate the optimal control for three systems with nonlinear dynamics (Fig.~\ref{f:systems}).
We find decompositions using Genetic Algorithm \cite{Khadke:2022} with and without modifying their system representation. From the discovered decompositions we compute policies using Policy Iteration \cite{Bertsekas:1995} and compare their closed-loop behavior.
Finally, we compare the decomposition policies to those obtained using the Advantage Actor Critic (A2C) \cite{Mnih:2016}, and the Proximal Policy Optimization (PPO) \cite{Schulman:2017} algorithms.

\subsection{Empirical Analysis with Linear Systems}
\label{subsec:linearemp}
We construct several optimal control problems for linear dynamical systems with different number of control inputs and state variables.
We apply two sampling strategies to randomly generate the dynamics and cost matrices
\begin{itemize}
    \item{\textbf{strategy I} : $\bs{A}$, $\bs{B}$, $\bs{Q}_s$ and $\bs{R}_s$ are sampled independently and uniformly over $[0,1]^{n\times n}$, $[0,1]^{n\times m}$, $[0,1]^{n\times n}$ and $[0,1]^{m\times m}$ respectively. $\bs{A}$ and $\bs{B}$ are the dynamics matrices, and $\bs{Q} = \bs{Q}_s\bs{Q}_s^T$ and $\bs{R} = \bs{R}_s\bs{R}_s^T$ are the cost matrices.}
    \item{\textbf{strategy II} : We sample a square optimal gain matrix $\bs{K}^\ast$ with equal singular values. Dynamics matrices $\bs{A}$ and $\bs{B}$, and cost matrices $\bs{Q}$ and $\bs{R}$ are chosen such that $\bs{K}^\ast$ is the solution to the LQR problem. Repeating singular values introduces additional degrees of freedom in the singular value decomposition. This sampling procedure is meant to test the regularization strategy (Appendix~\ref{apx:svd}) for resolving the non-uniqueness.}
\end{itemize}
For each of these, we compute mappings $\bs{T}_y$ and $\bs{T}_v$ using our approach described in Section~\ref{sec:transform} as well as their balanced realizations \cite{Moore:1981}, and evaluate the value-error (Eq.~\ref{e:ValueError}) for every possible decomposition of the original and transformed systems. 
\begin{table}[t!]
\caption{Number of linear systems that exhibit decompositions with lower value-errors when transformed with our approach described in Section~\ref{sec:transform}, and when transformed into a balanced realization \cite{Moore:1981}}
\label{tab:linearsystems}
\centering
\footnotesize
\begin{tabular}{c|c|c|c}
 &  & \multicolumn{2}{c}{$\bs{\#}$ \textbf{systems with lower}} \\
\textbf{strategy} & \textbf{system} &
\multicolumn{2}{c}{\textbf{value-error when transformed}}
\\[0.1cm]
\hhline{~~--}
& $(\#$ inputs, states) & ours & balanced realization\\
\hline
\parbox[t]{1mm}{\multirow{4}{*}{\textbf{I}}} & (2,4) & 97 / 100 & 56 / 100\\
& (2,6) & 100 / 100 & 48 / 100\\
& (2,10) & 99 / 100 & 45 / 100\\
& (2,15) & 100 / 100 & 44 / 100\\
 & (3,3) & 100 / 100 & 60 / 100\\
& (3,4) & 100 / 100 & 50 / 100\\
\hline
\parbox[t]{1mm}{\multirow{3}{*}{\textbf{II}}} & (2,2) & 92 / 100 & 27 / 100 \\ 
& (3,3) & 99 / 100 & 49 / 100\\
& (4,4) & 99 / 100 & 73 / 100\\
\end{tabular}
\end{table}
System representation derived using our method leads to decompositions with lower suboptimalities for $>90\%$ of the sampled linear systems, compared to $<75\%$ when transforming to a balanced realization (Table~\ref{tab:linearsystems}).

\subsection{Decompositions for Nonlinear Systems}
\label{subsec:bipedmanipquad}
\begin{figure}[h!]
    \centering
    \includegraphics[width=0.44\textwidth]{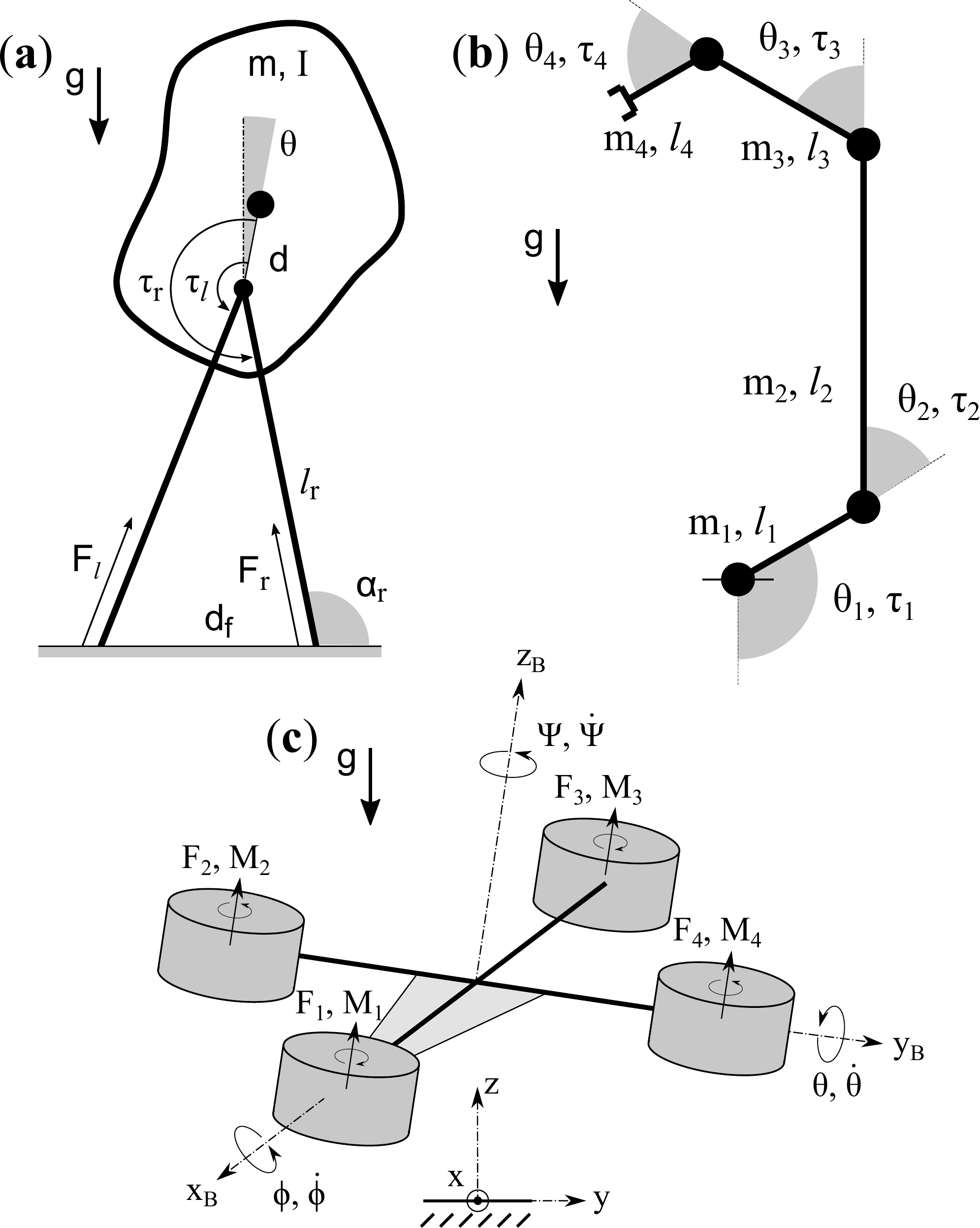}
    \caption{(\textbf{a}) Balance control of a simplified planar biped. (\textbf{b}) Swing-up control of a 4-link planar manipulator. (\textbf{c}) Hover control of a quadcopter.}
    \label{f:systems}
\end{figure}

We consider the problem of designing optimal controllers for the balance control of a simplified biped (Fig.~\ref{f:systems}(a)), the swing-up control of a planar manipulator (Fig.~\ref{f:systems}(b)) and the hover control of a quadcopter (Fig.~\ref{f:systems}(c)). The nonlinear dynamics are simulated in MATLAB.
The original system representations $(\bs{x}, \bs{u})$ used to describe these systems are,
\begin{itemize}
    \item{\textbf{Biped} : Four inputs (leg forces $F_{l/r}$ and hip torques $\tau_{l/r}$) and six state variables, the leg length $l_r$, leg angle $\alpha_r$, and torso angle $\theta$;  velocities, $\dot{x}$, $\dot{z}$, and $\dot{\theta}$. $(\bs{x}, \bs{u}) = ([l_r,\alpha_r,\dot{x},\dot{z},\theta,\dot{\theta}], [F_l, F_r, \tau_l, \tau_r])$.} 
    \item{\textbf{Manipulator} : Four inputs (joint torques $\tau_i$) and eight state variables, joint positions $\theta_i$ and joint velocities $\dot{\theta}_i$. $(\bs{x}, \bs{u}) = ([\theta_1,\theta_2,\theta_3,\theta_4,\dot{\theta}_1,\dot{\theta}_2,\dot{\theta}_3,\dot{\theta}_4], [\tau_1, \tau_2, \tau_3, \tau_4])$.}
    \item{\textbf{Quadcopter} : Four inputs (motor thrusts $F_i$) and ten state variables, centre-of-mass height $z$, attitude expressed in roll ($\phi$), pitch ($\theta$) and yaw ($\psi$), and velocities $\dot{x}$, $\dot{y}$, $\dot{z}$, $\dot{\phi}$, $\dot{\theta}$ and $\dot{\psi}$. $(\bs{x}, \bs{u}) = ([z,\phi,\theta,\psi,\dot{x},\dot{y},\dot{z},\dot{\phi},\dot{\theta},\dot{\psi}], [F_1, F_2, F_3, F_4])$.}
\end{itemize}

\begin{figure}
    \centering
    \includegraphics[width=0.485\textwidth]{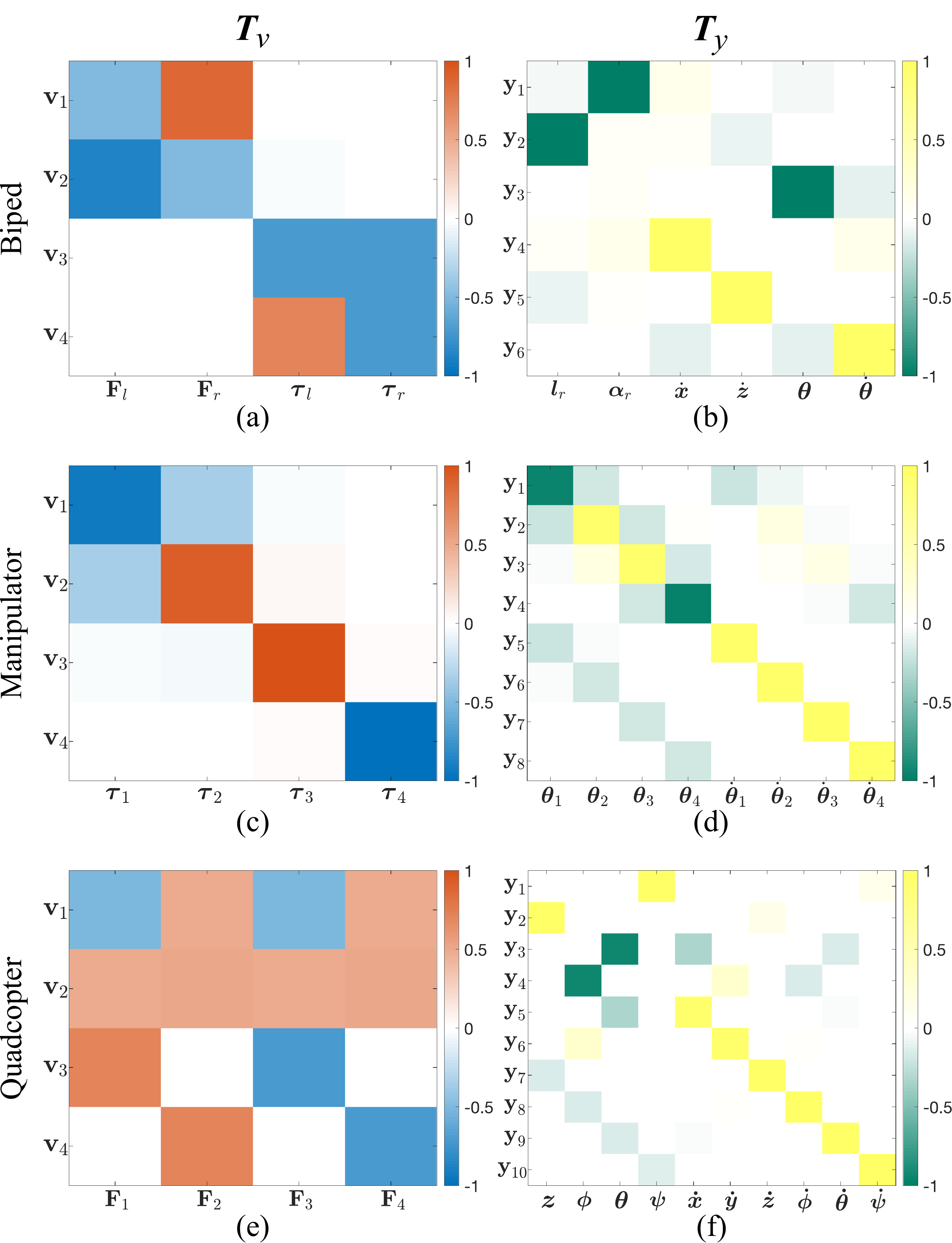}
    \caption{Mappings $\bs{T}_v$ and $\bs{T}_y$ for the systems shown in Fig.~\ref{f:systems}, derived using the approach presented in Section~\ref{sec:transform}}
    \label{f:transforms}
\end{figure}

We consider the linearized system dynamics at the goal state only to derive the linear mappings $\bs{T}_y$ and $\bs{T}_v$ using the approach in Section~\ref{sec:transform}; these are shown in Fig.~\ref{f:transforms}.
\begin{table*}[t!]
\caption{Comparison of the closed-loop performance of policies computed using the best decompositions for the original and transformed system representation with policies computed using A2C and PPO. 50 trajectories of the closed-loop system under the different policies are computed. Value function estimates are derived from costs of the trajectories that converge to the goal state, and normalized value function errors, as defined in Eq.~\ref{eq:normvalerr}, are reported in columns 2, 3 and 4. More negative the error, more optimal the decomposition $\delta_{(y, v)}$. Number of trajectories that converge to the goal are reported in column 5.}
\label{tab:decomptransformed}
\centering
\footnotesize
\begin{tabular}{P{1.2cm}|P{1.8cm}|P{1.8cm}|P{1.8cm}|P{3cm}}
\textbf{system} & \multicolumn{3}{c|}{\textbf{normalized value function error}} & $\bs{\#}$\textbf{trajectories converged} \\\hhline{~---~}
  & $\delta$ & A2C & PPO & $\delta$ / $\delta_{(y, v)}$ / A2C / PPO\\
\hline
\scriptsize{Biped} & $-0.29\pm0.49$  & $-0.39 \pm 0.61$ & $-0.08 \pm 0.35$ & 43 / 39 / 42 / 50 \\\hline
\scriptsize{Manipulator} & $-0.3\pm 0.58$ & $-1\pm 1.16$ & $-0.39 \pm 0.33$ & 50 / 50 / 38 / 49\\\hline
\scriptsize{Quadcopter} & - & $-1.14 \pm 1.34$ & $-1.47 \pm 1.25$ & 0 / 50 / 46 / 50\\
\end{tabular}
\end{table*}For the biped, the inputs are split into two groups in the transformed space, one consisting of the leg-forces $F_l$ and $F_r$, and the other the hip torques $\tau_l$ and $\tau_r$ (Fig.~\ref{f:transforms}(a)).
The transformed state variables $y_1$, $y_2$, $y_4$ and $y_5$ roughly describe the positions and velocities of the center-of-mass in the sagittal plane, whereas $y_3$ and $y_6$ correspond to the torso orientation and angular velocity respectively (Fig.~\ref{f:transforms}(b)).
For the manipulator, $v_3$ and $v_4$ correspond to the torques applied at joints 3 and 4 respectively, whereas $v_1$ and $v_2$ are a combination of torques at joints 1 and 2 (Fig.~\ref{f:transforms}(c)).
the state variable mapping is difficult to interpret (Fig.~\ref{f:transforms}(d)).
The mappings for the quadcopter are quite intuitive. Transformed inputs $v_2$, $v_1$, $v_3$ and $v_4$ are the scaled net thrust and differential yaw, pitch and roll thrusts respectively (Fig.~\ref{f:transforms}(e)).
Furthermore, transformed state variables can be split into four sets $\{y_1, y_{10}\}$, $\{y_2, y_7\}$, $\{y_3,y_5,y_9\}$ and $\{y_4,y_6,y_8\}$ corresponding to the system's yaw, altitude, pitch and roll descriptors (Fig.~\ref{f:transforms}(f)).

We use Genetic Algorithm \cite{Khadke:2022} to search for promising decompositions in the original and the transformed representation. The decomposition policies are represented as lookup tables over the state-space and computed using Policy Iteration \cite{Bertsekas:1995}. Additionally, as baselines to compare performance against, we compute policies using Proximal Policy Optimization (PPO) \cite{Schulman:2017}, and Advantage Actor Critic (A2C) \cite{Mnih:2016}. For PPO and A2C, we use the SB3 implementation \cite{Raffin:2021}, and the policies are feedforward neural networks with rectified linear unit activations. These neural network policies have two hidden layers of sizes $[128, 128]$, $[512, 512]$ and $[256, 256]$ for the biped, the manipulator and the quadcopter respectively. The PPO and A2C policies are trained for 20 million steps. At the end of training, the policy parameters (neural network weights) corresponding to the best performing policy are used for comparison.

To compare the closed-loop performance, we compute 50 trajectories for each system starting from different initial states with the different policies and note the number of trajectories that converge to the goal state (Table~\ref{tab:decomptransformed}, column 5). Note that, for the quadcopter, decomposing policy computation in the original system representation leads to unstable closed-loop behavior (none of the trajectories converge to the goal) whereas the policy computation in the transformed representation is readily decomposable and results in a stable closed-loop system (all trajectories converge to the goal). Furthermore, we estimate the value function for the different policies using costs of the trajectories that converge to the goal state. We use the normalized value function error,
\begin{equation}
    \big(\bs{V}^{\bs{\delta}_{(\bs{y},\bs{v})}} - \bs{V}\big) / \bs{V}^{\bs{\delta}_{(\bs{y},\bs{v})}}\label{eq:normvalerr}
\end{equation}
to quantify the relative quality of a policy, with value function estimate $\bs{V}$, in comparison to the decomposition policy derived from the transformed system representation, whose value function estimate we 
term $\bs{V}^{\bs{\delta}_{(\bs{y},\bs{v})}}$. More negative the error, relatively more optimal the decomposition policy. As can be seen in Table~\ref{tab:decomptransformed}, columns 2, 3 and 4, for all three systems the policy derived by decomposing the transformed system representation is more optimal than the one obtained from the original representation as well as the ones computed using A2C and PPO.

\section{Conclusions and Future Works}
\label{sec:conclusion}
We presented a heuristic to automatically discover system representations that are more suited for policy decomposition. 
Our approach constructs linear state and input mappings from a known system representation to a representation more amenable to decomposition. These linear mappings are derived from the singular value decomposition of the LQR approximation of the optimal policy. The singular value decomposition is not necessarily unique and we present a regularization strategy to resolve this non-uniqueness. For systems with linear dynamics our strategy almost always ($> 90\%$) produces representations that lead to decompositions with lower value-errors (Eq.~\ref{e:ValueError}).
For the optimal control of three nonlinear dynamical systems, viz a simplified biped, a planar manipulator and a quadcopter, decompositions discovered with the transformed system representation result in reduced trajectory costs. Moreover, the decomposition policies, represented as lookup tables over the state-space, produce trajectories with substantially lower costs compared to neural network policies derived from A2C and PPO. Although, representing policies as lookup tables becomes difficult with increasing dimensionality, PoDec finds promising reductions while sacrificing minimally on closed-loop performance.

There are several research directions to build upon this work. Firstly, further investigation is necessary to characterize the linear systems that lead to higher value-errors on transformation. Secondly, the search for system representations that induce sparsity in the optimal policies can be posed as a bi-level optimization problem
\begin{equation}
    \argmin_{\bs{T}_y, \bs{T}_v} \;\lVert \argmin_{\bs{\Theta}} V^{\pi_{\bs{\Theta}}}\rVert_1\nonumber
\end{equation}
In the above formulation, different trajectory optimization methods can be used to approximately solve the inner minimization. Moreover, a gradient based optimizer can be used to solve the outer minimization where $\bs{T}_y$ and $\bs{T}_v$ need not be restricted to linear mappings. Lastly, elucidating the connection between sparsity in the optimal policy and the minimum achievable decomposition value-error will help in rigorously formalizing the system representation search.



\section{APPENDIX}

\subsection{Resolving the non-uniqueness of SVD}
\label{apx:svd}
A singular value decomposition of $\bs{K}^\ast$ is $\bs{U}_{\bs{K}^\ast}\bs{S}_{\bs{K}^\ast}\bs{V}_{\bs{K}^\ast}^T$ where $\bs{U}_{\bs{K}^\ast}\in\mathbb{R}^{m\times m}$ and $\bs{V}_{\bs{K}^\ast}\in\mathbb{R}^{n\times n}$ are orthogonal, and $\bs{S}_{\bs{K}^\ast}\in\mathbb{R}^{m\times n}$ is the singular value matrix with singular values in a decreasing order along the main diagonal. Based on the singular values of $\bs{K}^\ast$ there are three possibilities
\begin{itemize}
    \item{\textbf{\textit{Case I}} : If all singular values are unique and non-zero then $\bs{U}_{\bs{K}^\ast}$ and the first $m$ columns of $\bs{V}_{\bs{K}^\ast}$ are uniquely defined. Last $(n-m)$ columns of $\bs{V}_{\bs{K}^\ast}$ span the null-space of $\bs{K}^\ast$, and we derive them by minimizing their L1 norm while constraining them to be mutually orthogonal as well as orthogonal to the first $m$ columns.}
    \item{\textbf{\textit{Case II}} : For the case of zero singular values, we jointly compute the columns of $\bs{V}_{\bs{K}^\ast}$ associated with zero singular values with the last $(n-m)$ columns as we did in \textbf{\textit{Case I}}. Corresponding columns of $\bs{U}_{\bs{K}^\ast}$ are also obtained by minimizing their L1 norm while restricted to being mutually orthogonal and orthogonal to the remaining columns of $\bs{U}_{\bs{K}^\ast}$.}
    \item{\textbf{\textit{Case III}} : Subsets of columns of $\bs{U}_{\bs{K}^\ast}$ corresponding to repeated and non-zero singular values span orthogonal subspaces (of dimension $> 1$) in the range of $\bs{K}^\ast$. Each of these subsets are derived by minimizing the L1 norm such that they are mutually orthogonal and orthogonal to the subspaces spanned by the remaining columns. Note that the orthogonality constraints between columns corresponding to different subsets are tied to the subspace they span rather than the choice of the columns (basis) themselves. Therefore, these subsets can be solved for in tandem. Consequently, the first $m$ columns of $\bs{V}_{\bs{K}^\ast}$ are completely determined from $\bs{K}^\ast$, $\bs{U}_{\bs{K}^\ast}$ and $\bs{S}_{\bs{K}^\ast}$. Last $(n-m)$ columns of $\bs{V}_{\bs{K}^\ast}$ are found as in \textbf{\textit{Case I}}.
    }
\end{itemize}

The L1 norm minimization for a mutually orthogonal set of column vectors $\bs{X}^i$, also orthogonal to a fixed subspace spanned by $\bs{Y}^k$ of the same dimension is posed as follows
\begin{equation}
    \begin{split}
&\min_{\bs{X}^i}\;\; \sum_{i=1}^{n_{X}}\Big(\lVert \bs{X}^i\rVert_1 + \lambda\sum_{k=1}^{n_{Y}}((\bs{X}^i)^T\bs{Y}^k)^2\Big)\\\text{s.t.}\; &(\bs{X}^i)^T\bs{X}^j = \bs{1}_{i==j}\; i,j\in\{1,\cdots,n_{X}\}
    \end{split}\nonumber
\end{equation}
where $\lambda$ is a fixed Lagrange multiplier (we use $\lambda=2000$) that incorporates the linear orthogonality constraint with subspace spanned by $\bs{Y}^k$. We use the approach in \cite{Wen:2013} to enforce the mutual orthogonality constraints and solve the optimization problem. A feasible initialization is obtained using the SVD routine in MATLAB \cite{Kahan:1990}. Note that, this optimization is not guaranteed to converge to the globally optimal solution but we found that transformations discovered still lead to representations with reduced optimalities.

\bibliographystyle{IEEEtran}
\bibliography{bibfile}

\end{document}